\documentclass[9pt,twocolumn,twoside]{pnas-new}
\usepackage{lineno}

\setboolean{displaywatermark}{false}
\templatetype{pnasresearcharticle} 

\begin{document}
\nolinenumbers
\title{Slowing translation to avoid ribosome population extinction and maintain stable allocation at slow growth rates}
\author[a]{Dotan Goberman}
\author[b]{Anjan Roy}
\author[a,1]{Rami Pugatch}

\affil[a]{Department of Industrial Engineering and Management, Ben-Gurion University of the Negev, Beer-Sheva 8410501, Israel}
\affil[b]{Department of Biochemical Engineering and Biotechnology
Indian Institute of Technology Delhi
Hauz Khas, New Delhi-110016, India}
\leadauthor{Goberman}

\significancestatement{Ribosomes in bacterial cells follow the ribosome growth law, which equates ribosome mass fraction times the ribosome translation speed with bacterial growth rate up to a constant. Previously, this constant was related to the ribosome lifetime and translation speed. Here, we show that at slow growth rates, close to the zero growth limit, the finite lifetime of ribosomes and their limited number present a strong constraint on allocating ribosomes to ribosomal protein synthesis. To maintain slow growth while avoiding the risk of extinction of the entire cellular ribosome population and loss of controllability over ribosome allocation, we show that some bacteria must decrease the translation speed significantly below its maximum, as observed.}

\authordeclaration{The authors declare no competing interests.}
\correspondingauthor{\textsuperscript{1}To whom correspondence should be addressed. E-mail: rami.pugatch@gmail.com}


\begin{abstract}
To double the cellular population of ribosomes, a fraction of the active ribosomes is allocated to synthesize ribosomal proteins. Subsequently, these ribosomal proteins enter the ribosome self-assembly process, synthesizing new ribosomes and forming the well-known ribosome autocatalytic subcycle. Neglecting ribosome lifetime and the duration of the self-assembly process, the doubling rate of all cellular biomass can be equated with the fraction of ribosomes allocated to synthesize an essential ribosomal protein times its synthesis rate. However, ribosomes have a finite lifetime, and the assembly process has a finite duration. Furthermore, the number of ribosomes is known to decrease with slow growth rates. The finite lifetime of ribosomes and the decline in their numbers present a challenge in sustaining slow growth solely through controlling the allocation of ribosomes to synthesize more ribosomal proteins. When the number of ribosomes allocated per mRNA of an essential ribosomal protein is approximately one, the resulting fluctuations in the production rate of new ribosomes increase, causing a potential risk that the actual production rate will fall below the ribosome death rate. Thus, in this regime, a significant risk of extinction of the ribosome population emerges. To mitigate this risk, we suggest that the ribosome translation speed is used as an alternative control parameter, which facilitates the maintenance of slow growth rates with a larger ribosome pool. We clarify the observed reduction in translation speed at harsh environments in E. \textit{coli} and C. \textit{Glutamicum}, explore other mitigation strategies and suggest additional falsifiable predictions of our model.
\end{abstract}

\dates{This manuscript was compiled on \today}
\doi{\url{www.pnas.org/cgi/doi/10.1073/pnas.XXXXXXXXXX}}

\maketitle
\thispagestyle{firststyle}
\ifthenelse{\boolean{shortarticle}}{\ifthenelse{\boolean{singlecolumn}}{\abscontentformatted}{\abscontent}}{}

\firstpage{3}

\dropcap{A}ll bacterial cells share a universal set of molecular machines that cooperatively self-replicate and are organized in a unique autocatalytic network architecture envisaged by von Neumann \cite{VN1,VN2,mypnas,ourPNAS}, not to be confused with the von Neuman architecture for a stored-program computer \cite{vnc}. A central part of this von Neumann autocatalytic network architecture for self-replication is the universal constructor---a machine that can synthesize all other machines, including itself, given raw material, energy, and a proper set of instructions. 

In the cell, the universal constructor is not comprised of a single machine but rather from a sub-network of essential molecular machines, better known as the transcription-translation machinery. The transcription-translation machinery can synthesize all the proteins in the cell and self-replicate its components \cite{ourPNAS}. Previously, we have shown that for any essential molecular machine in the cell, one can write a 'growth law' that connects the overall doubling time with the physiological parameters that characterize the machine---rate, allocation toward self-replication, and lifetime. In particular, we have shown that the well-known ribosome growth law \cite{Terry1,Terry2,Terry3,Terry4} can be identified as a private case of a growth law derived from the von Neuman architecture model. 

In \cite{CoryGlu,TranslationSpeedReduction}, it was experimentally demonstrated that ribosomes slow their translation speed at slow growth rates. In the same experiments, the RNA-protein ratio was shown to decrease approximately linearly as a function of the growth rate, but it approached a finite value at zero growth rate. To theoretically account for these experimental observations, we show that it is essential to account for two additional physiological processes: the ribosome assembly time---previously assumed negligible because it is much smaller compared to the doubling time, and the ribosome lifetime---previously neglected because it was much longer compared to the doubling times that were under consideration. 

Accounting for these two processes allows us to obtain a good fit for the measured data while providing a falsifiable prediction for the ribosome assembly time or its scaleless equivalent, the fraction of ribosomes in assembly, which we define explicitly in what follows, and for the ribosome lifetime. However, this leaves the question of why and under what circumstances it is evolutionarily beneficial to reduce the translation speed as the growth rate becomes smaller. This is because the same observed growth rates can be obtained with higher translation speeds and an associated reduction in ribosome allocation toward synthesizing ribosomal proteins.

Under slow growth conditions, the total number of ribosomes is known to decrease approximately quadratically with growth rate \cite{RMilo}. In E. coli, growing at a doubling time of $100$ minutes, the number of ribosomes is decreased $12$-fold compared to a doubling time of $20$ minutes. In the autocatalytic cycle of ribosomes---' ribosomes make more ribosomes,' a fraction $\alpha$ of the ribosomes is allocated to make ribosomal proteins, which enter the ribosome self-assembly process and synthesize new ribosomes. A fraction of the newly synthesized ribosomes is subsequently allocated to synthesize ribosomal proteins, thus completing the cycle. If the translation speed and duration of the ribosome assembly process are fixed, the only remaining way to control the growth rate is through the allocation parameter---$\alpha$. 

In the limit of long doubling times, of a few hundred minutes, the total number of ribosomes in E. \textit{coli} approaches $5400$ ribosomes (see SI). At such long doubling times, it becomes increasingly difficult to fine-tune ribosome allocation. For example, if the allocation parameter is $1 \%$, $1$ out of $100$ ribosomes are assigned to synthesize ribosomal proteins. In other words, on average there is one ribosome assigned to synthesize each ribosomal protein. Thus, a fluctuation in the allocation of a single ribosome can lead to large fluctuations in the synthesis rate of that ribosomal protein.

Additionally, as the ribosome synthesis rate approaches the ribosome death rate $\Gamma_D$ from above, fluctuations in the actual ribosome synthesis rate can result in underproduction and, as a result, lead to the extinction of the entire cellular ribosome population. 

Here, we show, that steady slow growth poses a new challenge to the bacterial cell, which is not present under fast growth conditions. In \cite{ProtDeath}, it was shown that protein degradation cannot be overlooked at slow growth rates, as it requires an increasing number of ribosomes to be allocated to maintain the proteome as the growth rate slows down. Here, in addition, we illuminate another aspect of the challenge of maintaining steady, slow growth. As the number of ribosomes allocated to make ribosomal proteins becomes of the order of one per essential ribosomal protein, temporal fluctuations in its synthesis below the ribosome death rate can potentially cascade into the extinction of the entire ribosomal population in the cell, as shown below. The cell utilizes the translation speed as an additional control parameter while fixing the allocation parameter to prevent the risk of cellular ribosome population extinction.
\subsection*{Results}
Using our generalized growth law, we extract the ribosomal lifetime and the fraction of ribosomes in assembly using the measured growth rate as a function of the RNA-to-Protein ratio and the measured translation speed (see Methods). 
Based on the measured translation speed as a function of growth rate, the known mRNA length of all ribosomal proteins, and the measured RNA-to-protein ratio, we used our theoretical model to infer the ribosome lifetime to be $500 \pm 60 (1\sigma)$ minutes (see Methods). 
The fraction of ribosomes in assembly---$\phi$, predicted from the model is $10 \%$---based on a calculation from one of the longest ribosomal protein operons (rplN-rplX-rplE-rpsN-rpsH-rplF-rplR-rpsE-rpmD-rplO-secY-rpmJ). This estimate changes up to $21 \%$ when it is based on another long ribosomal protein operon (rpsJ-rplC-rplD-rplW-rplB-rpsS-rplV-rpsC-rplP-rpmC-rpsQ), see Methods. Basing the calculation on all the large subunit ribosomal proteins also yields $\phi_i=20 \%$. In contrast, a calculation based on the rps ribosomal proteins or the rpm ribosomal proteins yields a higher estimate ($\phi_i=30 \%$ for rps proteins and $\phi_i=60 \%$ for rpm proteins), indicating that they are not limiting, as is believed to be the case. 

\begin{figure}[ht]
\centering
\includegraphics[width=.99\linewidth]{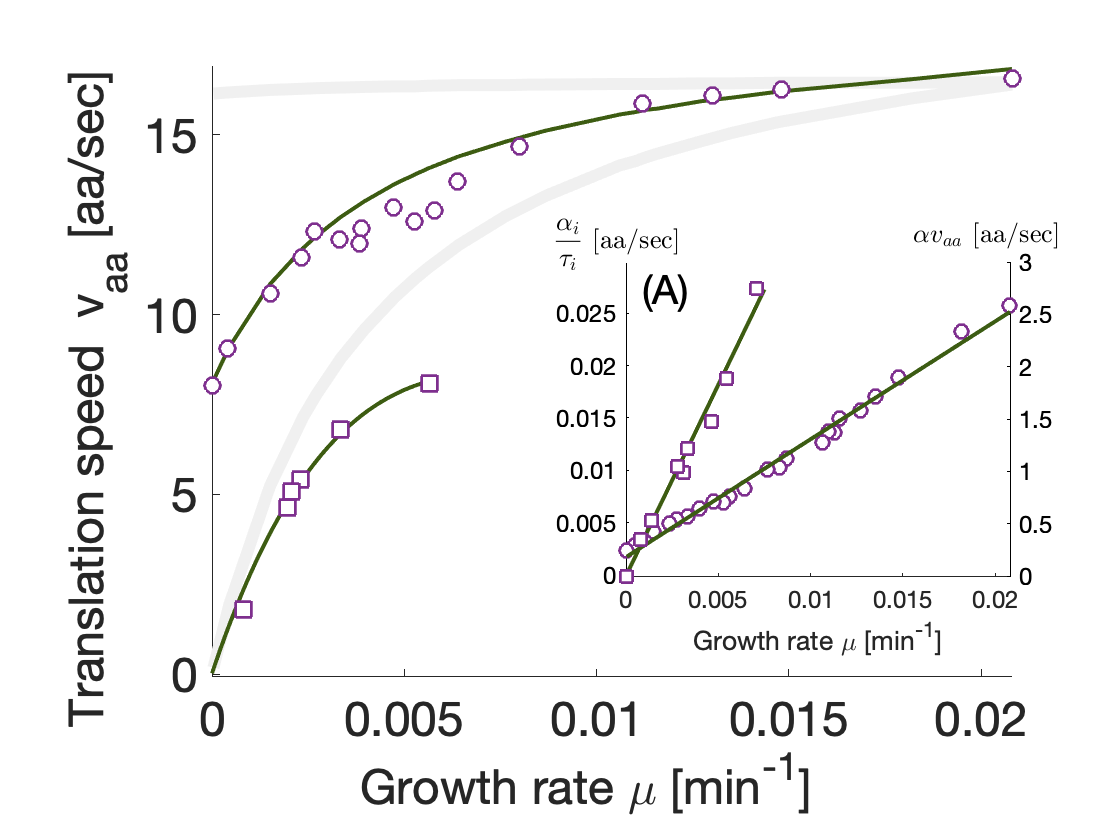}
\caption{Ribosome translation speed $v_{aa}$ in amino acids per second vs. growth rate in $[min^{-1}]$ for E. \textit{coli}, from \cite{Jun1} (purple circles) and \textit{Corynebacterium glutamicum} (purple squares) from \cite{CoryGlu} Theoretical curves are in green solid lines. The lower light gray solid line represents the limit of infinite ribosome lifetime. In this limit, we adjusted the ribosome assembly time to obtain a good fit for the translation speed at fast growth rates. The upper solid gray line represents the zero ribosome assembly time limit, where we adjusted the ribosome lifetime to obtain a good fit for the translation speed at fast growth rates. Both parameters are required to obtain a good fit for the data while accounting for the translation speed and the RNA-to-protein ratio as a function of the growth rate. Specifically, a finite ribosome lifetime is essential for explaining the observed nonzero RNA-to-protein ratio at zero growth rate. Accounting for both time scales, we obtain a prediction for the ribosome lifetime of E. \textit{coli} $\Gamma_D^{-1}=500 \pm 60$ min. Our model predicts a ribosome assembly time that varies linearly with the doubling time and approaches $150$ minutes at the zero growth rate limit, in accordance with \cite{williamsonScience}. In inset A, we show the data and the fit from which we extracted the ribosome assembly time and lifetime. The left y-axis and the corresponding purple circles depict data for E. \textit{coli} taken from \cite{Jun1}, and the right y-axis and the corresponding purple squares depict data for \textit{Corynebacterium glutamicum}, taken from \cite{CoryGlu}. A linear green solid line depicts our theory. It fits well with the data, using two fitting parameters as described in the main text, namely $\phi$, which can be related to the assembly duration, via the growth law \ref{eq:two}, and $\Gamma_D$, which, is the ribosome death rate.}
\label{fig:one}
\end{figure}

Utilizing these two parameters, namely $\phi_i=0.1, \Gamma_D = 500$ $min^{-1}$ enables us to create a theoretical curve for the ribosome translation speed (measured in amino acids per second) as a function of the growth rate; this curve nicely fits the measured data from reference \cite{TranslationSpeedReduction}, As seen in Fig. \ref{fig:one}. 

The two fitted parameters of our model, namely the ribosome lifetime and the ribosome fraction in assembly, can be measured and contrasted with our predictions. 
Using these parameters, we also obtained a good fit for the measured dependence of the RNA-to-protein ratio as a function of the growth rate. 

We performed a similar analysis for another type of bacterium, \textit{Corynebacterium Glutamicum}, a gram-positive bacteria belonging to the actinomycetes class \cite{CoryGlu}. The results, which are consistent with our model, are presented in Fig. \ref{fig:one} and in inset A of Fig. \ref{fig:one} (data depicted by squares, theory in solid green). Since we did not have data on $\alpha_i$ we plotted $\alpha v_{aa}$ as a function of the growth rate $\mu$, where $v_{aa}$ is the ribosome translation speed measured in amino acids per second. We found that $\alpha v_{aa}$ is a linear function of the growth rate $\mu$, in accordance with the prediction of our theory.  

\begin{figure}[ht]
\centering
\includegraphics[width=1\linewidth]{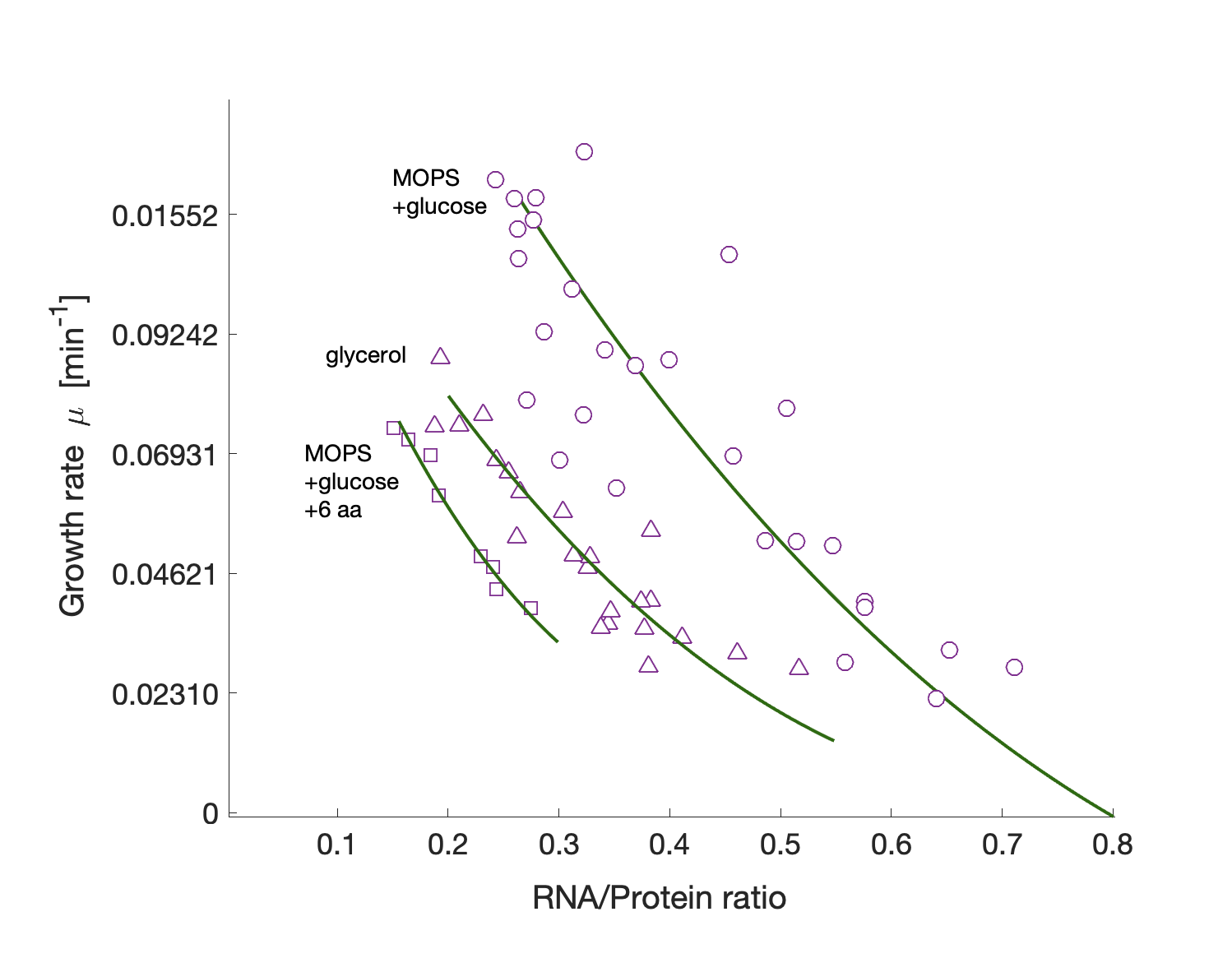}
\caption{Growth rate vs. RNA-to-protein ratio. Depicted in open squares, triangles, and circles are data from E. \textit{coli} grown in 3-morpholinopropanesulfonic acid (MOPS) buffer environments with glucose and with a supplement of six amino acids (squares) or without (circles). The triangles represent an environment where the carbon source comes from glycerol; see \cite{Jun1}. In all these environments, varying concentrations of antibiotics were administered. The data depicted by open triangles (i.e., the glycerol environment) is from an experiment that administered variable amounts of erythromycin antibiotic drug. The other data points (depicted by circles and squares) are from experiments that used chloramphenicol antibiotic drug \cite{Jun1}. To obtain the fits depicted by green solid lines, we modulated the ribosome death rate according to a Michaelis Menten term with a single fitting parameter, yielding a fair fit to the data.}
\label{fig:two}
\end{figure}

Since ribosome lifetime is a parameter that enters our model, we wanted to test it against experimental data where ribosome lifetime is manipulated. For that purpose, we used data from \cite{Jun1} on the dependence of growth rate on the concentration of antibiotics that specifically target the ribosome and hence affect its lifetime. 
In \cite{Jun1}, bacteria were grown in different media with varying concentrations of antibiotics. We focused on chloramphenicol and erythromycin, both antibiotics that target the ribosome. We use our version of the growth law, Eq. \ref{eq:two}, and assume that the concentration of antibiotics adversely modulates the lifetime of ribosomes via a Michaelis-Menten-like dependence: $\bar{\Gamma}_D^{-1}(c)=\Gamma_D^{-1}\frac{k}{k+c},$
where $\bar{\Gamma}_D^{-1}(c)$ is the ribosome lifetime as a function of the concentration of antibiotics $c$, $\Gamma_D^{-1}$ is the ribosome lifetime without antibiotics which we found before, and $k$ is the fitted Michaelis-Menten saturation parameter. The good fit obtained in Figure \ref{fig:two} further consolidates our model.

Our extended ribosome growth law, Eq. \ref{eq:two}, accounts for the relationship between the ribosome translation speed, the RNA-to-protein ratio, and the growth rate by using the ribosome lifetime and the ribosome assembly duration as additional fitting parameters to the standard ribosome growth law \cite{Terry1,Terry2,Terry3,Terry4}. 

However, we can also ask why the ribosome translation speed becomes slower at slow growth rates since this is not the only possibility consistent with the observed growth rate. For example, the same growth rate could have been obtained by reducing the allocation parameter while keeping the translation speed at its maximum. This is what happens in other bacteria and even in E. coli at fast growth rates. 

In E. \textit{coli}, the absolute number of ribosomes decreases approximately quadratically as a function of the growth rate (SI, data taken from \cite{RMilo}). 
If the only way to control the growth rate was via ribosome allocation, the translation speed would have been kept at its maximal value by keeping the charged tRNA and elongation factor G (EF-G) pools fully charged. In such a scenario, the fraction of ribosomes allocated to make ribosomal proteins would decrease as the growth rate decreases. However, at some point, the number of ribosomes dedicated to synthesizing each one of the essential ribosomal proteins becomes of the order of one, i.e., just a few ribosomes per protein. At this point, any random fluctuation in the allocation can lead to large fluctuations in the synthesis of new ribosomes. If these fluctuations cause the ribosome production rate to go below the ribosome death rate, the ribosome population will decrease, and there is a risk that it will become extinct. We thus see that, perhaps unexpectedly, there is a control challenge in sustaining slow growth rates.

\subsection*{Mitigating ribosome population extinction risk} A well-known result from branching processes \cite{branching} states that if the doubling time of a population of $N$ self-replicating agents is shorter than their lifetime, the extinction probability tends to zero at the large $N$ limit. However, as we show, if the number $N$ is small, the extinction probability can become non-negligible. 

When the number of ribosomes allocated to synthesize a specific essential ribosomal protein in the cell is of the order of one, we enter a regime known as the shot noise limit. In this regime, the instantaneous synthesis rate of new ribosomes, which directly depends on manufacturing critical components, i.e., synthesizing essential ribosomal proteins, significantly increases. These fluctuations can cause the synthesis rate of new ribosomes to drop below the ribosome death rate, consequently leading to ribosome population extinction.   

While this risk can be mitigated by reallocating more ribosomes, as seems to be the case when antibiotics that attack the ribosomes are presented, this can be very costly to the cell, especially if these ribosomes are required elsewhere \cite{KimatVeNifanu}. We argue that from an evolutionary perspective, other mitigation strategies might come into play. One such strategy is to slow down ribosome translation speed.

\begin{figure}[ht]
\centering
\includegraphics[width=.8\linewidth]{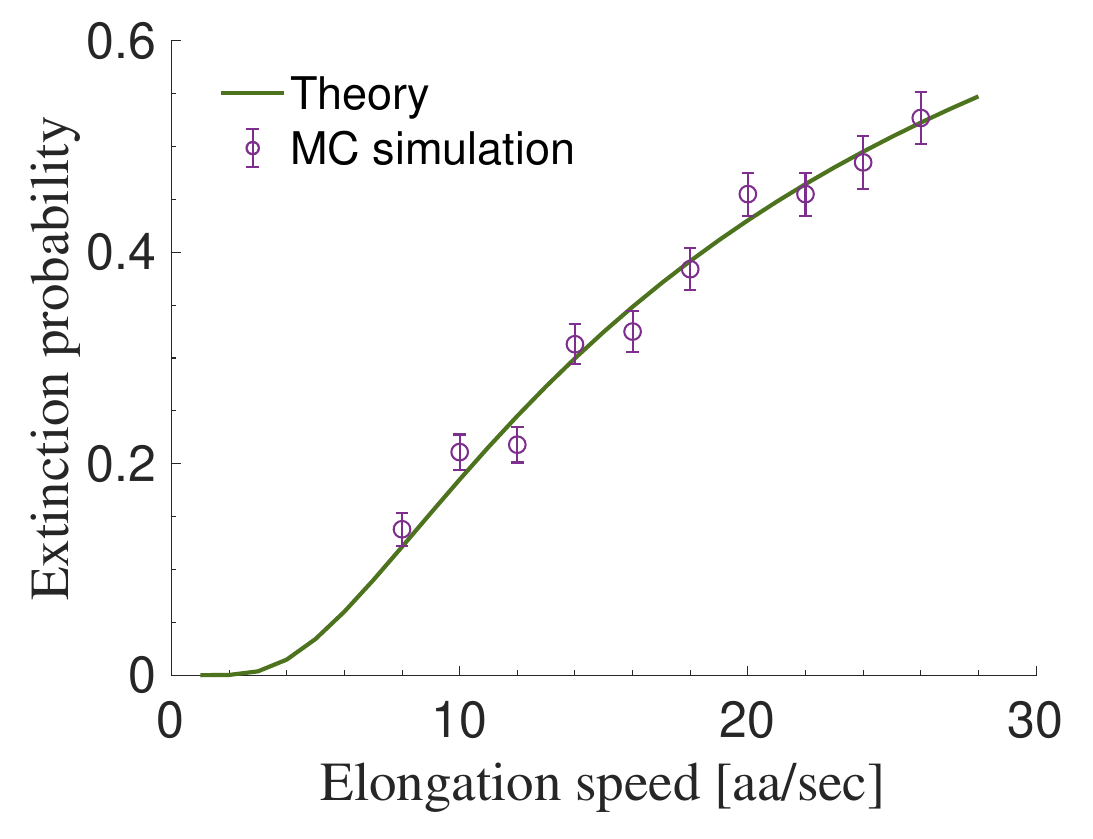}
\caption{Extinction probability for a fixed average doubling time of $160 \pm 25 (1 \sigma)$ min. The green solid curve represents the theoretical calculation of the extinction rate based on Eq. \ref{eq:extinction5}, and the circles are the result of a Monte-Carlo simulation of the ribosome autocatalytic cycle, with a random assembly time whose average is $\approx 30$ min. Even though the theoretical model is simpler and less detailed compared to the simulation, it fits the simulation results well.}
\label{fig:three}
\end{figure}

To demonstrate the effect of this strategy on the ribosome population extinction probability, we calculated the probability of extinction with a given fixed growth rate and a fixed initial number of ribosomes as a function of the ribosome translation speed. We present our results in Figure \ref{fig:three}. Using a Monte-Carlo simulation of a population of ribosomes engaged in the sub-cycle that self-assemble more ribosomes, specifically focusing on the large subunit, which is thought to be the bottleneck for ribogenesis, we estimated the extinction probability from the simulation, by directly counting extinction events in the simulation. Mean values are plotted as circles in figure \ref{fig:three} while error bars depict the standard deviation. We also plot the extinction probability, which we calculated theoretically (solid line); see Methods for the derivation of the analytical expression.  

As the figure shows, the risk of ribosome population extinction increases monotonically with the translation speed for a given growth rate ($160$ minutes doubling time in the figure). For example, slowing down ribosome speed from its maximal rate (22 aa/sec) to a rate of 10 aa/sec decreases the extinction probability by a factor of three.

We want to emphasize the non-trivial aspect of the accordance between the simulation and the theoretical calculation. While the simulation accounted for the random parallel synthesis of all the ribosomal proteins and their assembly process, and specifically for the fact that a new ribosome cannot form before all its sub-components are synthesized, the extinction probability calculation used a simple model for the extinction of a finite population of replicating agents without any inner structure. The accordance is thus a sign that this approach toward estimating the extinction rate is valid and can also be used for biological cells. 

The results presented in Fig. \ref{fig:three} consolidate our hypothesis about the evolutionary logic behind translation speed reduction in harsh environments. Finally, we note that the analytical calculation of the extinction probability of a finite population of $N$ replicating agents is not present in the standard literature that we are aware of in branching processes, e.g., \cite{branching}, perhaps because branching processes results are often derived in the limit $N \gg 1$. See however \cite{OrnP} for a calculation of the dynamics towards extinction (with probability one), albeit with correlated death of both mother and daughter.  

\subsection*{Discussion} In industrial engineering, adjusting production to meet demand when both fluctuate randomly requires buffers. There are three types of buffers \cite{FactoryPhysics}; (i) a time buffer, which allocates extra time to a bottleneck process; (ii) an inventory buffer, which retains raw material, partially manufactured goods, or finished goods in excess for usage whenever demand increases beyond the momentary ability to satisfy it via production; (iii) a capacity buffer, which allows machines along the production line to increase their speed assuming they are not operating at their maximum speed. 

A cell can be viewed as a complex factory whose product is another factory---an autocatalytic network \cite{ourPNAS}. As discussed, all self-reproducing cells have a unique autocatalytic network structure known as the von Neuman architecture.  
In a given environment, a cell can maximize its growth rate by finding an internal allocation of its machines, particularly of the machine that makes all the machines---the universal constructors (UCs). The allocation parameter $\alpha$ describes the fraction of universal constructors dedicated to manufacturing more universal constructors. The remaining fraction of universal constructors $1-\alpha$ is allocated to construct all the other machines that are required to maintain the entire operation, e.g., to supply the universal constructors with raw material, energy, instructions, an environment to function within, i.e., a membrane that creates a difference between an internal and an external environment. Note that some of the UCs can be allocated to be inactive, e.g., by modulating the initiation of translation. Interestingly, as shown in \cite{Wingreen}, finding the optimal allocation is readily obtained via a product feedback mechanism.   

In principle, all three types of buffers can be utilized to control the manufacturing rate and adjust it to the external, possibly random, supply. In a cellular context, the inventory buffer probably received more attention. Inventory buffers in metabolism known as metabolic pools \cite{MetabolicPool} enable the streamlined production of essential metabolites, e.g., amino acids, nucleotides, and oligosaccharides, even when the production rate of individual metabolites fluctuates. As an additional example, RNA-polymerase is known to be in excess in E. coli, and this excess can be viewed as an inventory buffer, which, as suggested in \cite{KimatVeNifanu}, is required to mitigate noise in gene expression and facilitate growth. Free ribosomes can also be considered as an inventory buffer.

A capacity buffer occurs when control over the speed of molecular machines is exerted such that they are not working at their maximal rate under all circumstances. As seen in Figure \ref{fig:three}, using a capacity buffer facilitates a reduction in the extinction probability without affecting the growth rate; hence, at low growth rates, using hybrid control of both inventory and capacity is better than using just one of them. Interestingly, the elongation speed of RNA polymerase is controlled by the highly conserved protein spt4 \cite{spt4}. In \cite{TransRate}, a measurement of the elongation speed of RNA-polymerase demonstrated it decreases as the growth rate decreases. 

A time buffer can be relevant when the allocation parameter is controlled temporally. For example, a circadian clock can regulate gene transcription, resulting in scheduled protein production \cite{CyanoBact}. Suppose there is a delay in the production of essential proteins. In that case, the circadian clock can be delayed, e.g., by a feedback regulator, to allow the essential proteins to be synthesized and the shortage gap to be closed---a time buffer \cite{FactoryPhysics}.  
As presented in Methods, the growth law that takes into account both ribosome lifetime and the finite duration of the ribosome assembly process is given by 
\begin{eqnarray}
\mu = (\frac{\alpha_i}{\tau_i} - \Gamma_D) \frac{1}{1+\phi_i}.
\label{eq:two}
\end{eqnarray}

This equation can be interpreted from the point of view of buffer allocation. Modulating the allocation parameter in time using a predefined schedule and modifying this schedule if critical processes lag behind schedule is analogous to time-buffer control \cite{FactoryPhysics}. If the allocation is not over time but rather asynchronous, controlling the growth rate can be achieved by controlling the level of inventory of ribosomes-in-assembly via $\phi_i$'s, which is known to be locked to a constant value via a translational feedback loop first discovered by Nomura \cite{Nomura1,Nomura2}. Thus, we conclude that controlling the translation speed to control the growth rate is an example of using a capacity buffer in a cellular contest. 

Next, following \cite{ProtDeath}, we focus on the relation between the inactive fraction of ribosome and the growth rate. Without control over the inactive fraction, the inactive ribosomes diffuse "in between jobs" from one mRNA to another. Without specific control, e.g., on the initiation of translation, it is expected that as the allocation of ribosomes toward ribosomal protein synthesis is decreased, the allocation of ribosomes to synthesize other proteins will increase. To break this negative correlation, the inactive fraction must be under control, i.e., to serve as a buffer. At slow growth, as the ribosome allocation parameter $\alpha$ alongside the translation speed $v_{aa}$ decreases, the only way to prevent an unwarranted increase in the synthesis of other proteins is to increase the inactive fraction. While the exact mechanism behind this requires further study, the inactive fraction is experimentally observed to increase as the growth rate slows down \cite{ProtDeath}. 

As shown, in harsh environments, there is a limit to how small the allocation parameter $\alpha_i$ can be set since the finite number of ribosomes and their finite lifetime put a shot noise limit on how low we can safely allocate them without risking their extinction. Indeed, if we want to allocate a small percentage of ribosomes towards ribosomal protein synthesis, such that the average number of ribosomes allocated per essential ribosomal protein is of the order of one, random fluctuation in the actual number of ribosomes allocated can lead to a drop in the ribosome synthesis rate, below their death rate thus risking a cascade to total extinction of the entire ribosome population. 

An interesting alternative for controlling ribosome allocation, which we did not investigate in this paper, is to temporally allocate significant portions of the ribosome population to synthesize proteins in a predefined schedule. We postulate that in very small cells, e.g., SAR11 \cite{SAR11,SAR11b} or Prochlorococus Marinus \cite{proch}, the small number of ribosomes, $\sim 100$ in SAR11, forces the cell to employ this control mechanism---temporal ribosome buffer management, to sustain its growth.

In an autocatalytic network, all cycles share the same growth rate at steady-state.
In a bacterial cell, this observation \cite{ourPNAS} suggests that there exist two generic "control knobs" for the growth rate; (i) The allocation parameter $\alpha$ that controls what fraction of the transcription-translation machinery is dedicated to self-replicate itself; (ii) The translation speed of ribosomes alongside with transcription speed of RNA-polymerase \cite{Lagomarsino,ourPNAS} controlling how fast these central molecular machines drive autocatalysis. 

There is no a priori reason to believe that all autocatalytic cycles employ the same control strategies. However, at a low growth rate,  we claim that all cycles will face the shot noise limit, and to mitigate it, they will either employ the speed control knob or the temporal allocation control knob. For example, to maintain high enough levels of RNA-polymerase well above the extinction threshold, using similar arguments, we expect transcription speed to decrease in harsh environments when the growth rate is low. Indeed, in E. \textit{coli}, the transcription speed of RNA polymerase decreases with the growth rate as measured in \cite{TransRate}, see also \cite{RMilo}. 

The shot noise extinction threshold can also be found in other autocatalytic systems. A historically well-known, albeit dual example, comes from the Manhattan Project. When calculating the critical mass of U235, legend tells that Feynman noticed that as a subcritical mass approaches criticality from below, the risk that a random chain reaction will lead to a nuclear explosion increases significantly. Similarly, if the mass is above criticality but close to it, an initiated nuclear reaction can be extinct and die out without achieving conversion of most of the fissile material. 

In epidemiology, an epidemic outbreak with a reproductive ratio $R_0>1$ can be extinct due to stochasticity, before infecting a substantial percentage of the population \cite{FailR}. Conversely, an epidemic with $R_0<1$ can grow exponentially and infect a large percentage of the population before being extinct \cite{FailR}.  

The question of the minimal machinery required for sustaining life or the physiological determinants of spore viability is also related. We hypothesize that the stochastic nature of spore germination can be traced back to the small number of ribosomes and RNA-polymerase in the spore. Misallocating ribosomes, for example, in the early stages of the germination process, can cause premature extinction instead of growth. 

Interestingly, when the challenge is rebooting life, i.e., re-initiating growth after a period of stasis rather than sustaining slow growth, the order in which genes are transcribed and translated can significantly affect cell fate. In this regime, the small number of essential molecular machines, such as ribosomes, RNA-polymerases, metabolic proteins, and transporters, increases sensitivity to random fluctuations. 

For example, suppose that the cell initiates the synthesis of ribosomal proteins before expressing enough transporters and metabolic proteins. In that case, the cell can get trapped in a starved state, lacking the essential nutrients and ATP required to continue. However, a cell that first expresses metabolic transporters and metabolic proteins and then ribosomal proteins can escape this fate after a lag phase. In \cite{UriOrder}, it was shown that the expression of metabolic genes, which are part of a metabolic pathway, is matched to their order in the pathway. Based on our model, we predict that reversed order will not be essential for cell survival under fast growth conditions but can be critical when the growth rate is low and the number of ribosomes is small. We thus predict that reversing the order of expression can cause significant lag and lead to increased cell death, compared to the wild type, in harsh environments, not nutrient-rich environments.   

In \cite{ArcheaPaper}, a fixed number of ribosomes was measured across a wide range of growth rates. Based on our model, we predict that the translation speed of this single-cell organism will change linearly with the growth rate, in accordance with Eq. \ref{eq:four}. Based on our view regarding the evolutionary reason for choosing such a work point, we predict that this Archea has a small number of ribosomes, at least an order of magnitude less than E. \textit{coli}. We hypothesize that this work-point has been chosen to maintain a safe distance from the ribosome extinction threshold while allowing growth rate control via the translation rate.  

\section*{Methods}
Consider a simplified model for the ribosome autocatalytic sub-cycle \cite{ourPNAS}, in which ribosomes randomly create ribosomal proteins from ribosomal protein mRNA. The ribosomal proteins join the ribosome assembly process and bind to their target site directly on rRNA or in the partially assembled ribosome, eventually forming new ribosomes, thus completing the cycle. 

It is often implicitly assumed that the ribosome's death rate is negligible, i.e., the ribosome lifetime is significantly larger than the doubling time. Similarly, the ribosome assembly duration is also neglected, i.e., assumed to be very small compared to the doubling time. Instead, we take both of these time scales to be finite. 

We define $\phi_i$---relative work-in-process (relative WIP) as the relative number of ribosomal proteins of type $i$ that are free or part of a partially assembled ribosome relative to the total number of active ribosomes in the cell. For example, at a doubling time of $24$ minutes, it is estimated that, on average, an E. \textit{coli} will have $72000$ ribosomes \cite{RMilo}. Thus, if the relative WIP of a specific ribosomal protein is $10 \%$, there are $7200$ such proteins in different stages of the ribosome assembly process. 

Consider a deterministic process with a ribosome lifetime of $\tau_D$, translation duration $\tau_i$, and assembly duration $\tau_{SA}$ translate proteins. Let $U_0(t)$ be the number of ribosomes just about to be free. Then,
\begin{align}
U_0(t) = &U_0(t-\tau_i) + \alpha_i U_0(t-\tau_i-\tau_{SA_i}) \nonumber \\ &-\alpha_i U_0(t-\tau_i - \tau_{SA_i}-\tau_D) 
\label{eq:tmp1}
\end{align}
i.e., the number of ribosomes that are just freed is equal to the number of ribosomes that started translating $t-\tau_i$ minutes ago: $U_0(t-\tau_i)$, plus the number of ribosomes just born from the assembly process whose duration is $\tau_{SA_i}$ minutes: $\alpha_i U_0(t-\tau_i-\tau_{SA_i})$, minus the number of ribosomes that were born exactly a lifetime $\tau_D$ ago and are hence are dying just now: $\alpha_i U_0(t-\tau_i-\tau_{SA_i}-\tau_D)$.

Consider the ribosome self-assembly process, where ribosomal proteins enter the assembly process to bind with rRNA or with a partially assembled ribosome. Assume a ribosomal protein $a_i$ to be limiting. Then, it is taken to the assembly process as soon as it is synthesized. Let $\tau_{SA_i}$ be the residual assembly duration measured from when $a_i$ enters the ribosome assembly process to when the ribosome assembly is complete, with $a_i$ embedded in a ribosome. Finally, let the fraction of ribosomes allocated to translate protein $a_i$ be $\alpha_i$. Only free ribosomes can be allocated to do so, hence the number of newly formed ribosomes that are just born is $\alpha_i U_0(t-\tau_i-\tau_{SA_i})$. Similarly, one can identify the number of ribosomes that die right now as those born $\tau_{SA_i}-\tau_D$ minutes ago. 

Consider next the total number of ribosomes at time $t$: $U(t)$ (note that $U(t) \neq U_0(t)$). Then,
\begin{eqnarray}
U(t) = U(t-\tau_{SA_i}) + a_i(t-\tau_{SA_i}) -a_i(t-\tau_{SA_i}-\tau_D),
\label{eq:tmp2}
\end{eqnarray}
i.e., the total number of ribosomes right now equals the number of ribosomes that were present $t-\tau_{SA_i}$ minutes ago, plus the newly born ribosomes right now, $a_i(t-\tau_{SA_i})$ minus the ribosomes that were born a lifetime $\tau_D$ minutes ago, and hence are now about to die: $a_i(t-\tau_{SA_i}-\tau_D)$.

Next we insert an exponential ansatz: $\vec{v}(t)=\vec{v_0}e^{\mu t}$, in equation \ref{eq:tmp1}, where $\vec{v}(t)=(U(t),a(t))$, which holds at steady-growth conditions due to the autocatalytic nature of the process. Assuming all time scales are distributed randomly and independently from known distributions, we integrate over all times to obtain a simple growth law: 
\begin{eqnarray}
1 = P_{\tau_i}(\mu) \left( 1 +\alpha P_{\tau_{SA_i}}(\mu)(1-P_D(\mu)) \right),
\label{eq:lap2}
\end{eqnarray}
where $P_{\tau_i}(s=\mu)$ is the Laplace transform of the distribution of protein synthesis durations, $P_{\tau_{SA_i}}(s=\mu)$ is the Laplace transform of the (residual) distribution of ribosome assembly duration, and $P_D(s=\mu)$ is the Laplace transform of the distribution of ribosome lifetimes, all evaluated at growth rate $s=\mu$.

Next, we insert the same exponential ansatz: $\vec{v}(t)=\vec{v_0}e^{\mu t}$ in \ref{eq:tmp2} and obtain a second growth law for the growth rate $\mu$ as a function of the Laplace transform of the distribution of residual assembly times of protein $a_i$ and its relative WIP $\phi_i = \frac{a}{U}$:
\begin{eqnarray}
1=P_{\tau_{SA_i}}(\mu)\left(1+\phi_i\left( 1-P_D(\mu) \right) \right),
\label{eq:lap3}
\end{eqnarray}
Suppose that the distribution of residual assembly duration is exponential. In that case, Eq. \ref{eq:lap3} reduces to the following growth law of ribosome assembly: 
\begin{eqnarray}
\mu = \frac{\phi_i}{\tau_{SA_i}} - \Gamma_D
\label{eq:growthassembly}
\end{eqnarray}
i.e., the growth rate equals the rate of production of new ribosomes minus their death rate $\Gamma_D=\frac{1}{\tau_D}$. 

Similarly, if we further assume exponential distributions and use Eq. \ref{eq:lap2} we obtain:
\begin{eqnarray}
\mu = (\frac{\alpha_i}{\tau_i} - \Gamma_D) \frac{1}{1+\phi_i}
\label{eq:four}
\end{eqnarray}

Assuming $\phi_i$ equal zero and noting that $v_{aa} =\frac{L_i}{\tau_i}$ we obtain: 
\begin{eqnarray}
\mu = \frac{v_{aa}}{L_i}(\alpha_i - \frac{\Gamma_D L_i}{v_{aa}}).
\label{eq:four_terr}
\end{eqnarray}
Identifying $\kappa = \frac{v_{aa}}{L_i}$, $\Phi=\frac{\alpha_i}{L_i}$ and $\Phi_0=\frac{\Gamma_D}{v_{aa}}$  we obtain the known ribosome growth law  \cite{Terry1,Terry2,Terry3,Terry4}.
 
So far, we presented ribosome growth laws that are derived for a single essential ribosomal protein. This is by no means a constraint, and we can also obtain, after some algebra, growth laws that depend on sets of ribosomal proteins, including all of them in one set:
\begin{eqnarray}
\mu (L+ \hat L) = v_{aa}(\alpha - \frac{\Gamma_D}{v_{aa}}L),
\label{eq:four_group}
\end{eqnarray}
where $L$ is the total ribosomal protein length in amino acids and $\hat L$ is the sum over all the ribosomal protein lengths weighted by their relative part in the WIP (see SI for a more detailed example).

To obtain the theoretical curve depicted in Figure 1, we first used Eq. 4, namely, $(\frac{\alpha_i}{\tau_i})=\mu \times (1+\phi_i) + \Gamma_D$ and used both the measured RNA to protein ratio and the translation speed-$v_{aa}$ as a function of growth rate. 

First, we calculated the allocation parameter $\alpha$ from the RNA-to-protein ratio measurements ($\frac{R}{P}$) and knowledge of the ribosome rRNA content---which is $2/3$ of the ribosome's mass \cite{R2P2_3}. We equate $\alpha=0.9 \times \frac{1}{2} \frac{R}{P}$, where $0.9$ approximately accounts for the rRNA fraction from the total RNA measured \cite{RNA09}, see SI for further details. To obtain the allocation parameter per ribosomal protein $\alpha_i$, we assume optimal allocation, i.e., that there is a product feedback inhibition control that makes sure no ribosomal protein is overly synthesized relative to the others \cite{Nomura1,Nomura2}. 
This assumption leads to the following relation, $\frac{\alpha_i}{\alpha_j}=\frac{L_i}{L_j}$, where $L_i$ is the number of amino-acids in ribosomal protein $i$. Thus, $\alpha_i=\alpha \frac{L_i}{\sum_i L_i}$.  

We assume that the translation speed $v_{aa}$ is roughly the same across the mRNA of all ribosomal proteins. Thus, $\tau_i = \frac{L_i}{v_{aa}}$. Using these two relations, we plotted the allocation parameter times $v_{aa}$ as a function of the growth rate $\mu$ to obtain both the ribosome lifetime and the WIP: $\phi_i$, which are the slope (minus one) and the y-intercept of this graph (see data in inset A of Fig. \ref{fig:one}, depicted with open circles). 

Knowing $\phi_i$ and $\Gamma_D$, we calculated from our theory (Eq. 4) the translation speed $v_{aa}$ per growth rate and compared it to the measured values, as seen in Fig. 1. 

\subsection*{Calculating the probability of ribosome population extinction}
To calculate the extinction probability of a population of self-replicating agents, we first calculated the probability of extinction of a single replicator. Let $q^*$ be the probability a single replicator to die before replicating, i.e.,
\begin{eqnarray}
q^*=prob(\tilde{\tau}>\tilde{L}-\tilde{A}),
\label{eq:qstar}
\end{eqnarray}
where $\tilde{\tau}$ is the random doubling time, $\tilde{L}$---random lifetime, and $\tilde{A}$---random age, so $\tilde{L}-\tilde{A}$ is the random variable that represents the residual lifetime of the replicating agent. 
Let $q$ be the probability of extinction of a newborn ($\tilde{A}=0$) before replicating. Let $P^*$ be the probability of extinction of a population that started with a single replicating agent,
\begin{eqnarray}
P^*&=&q^*+(1-q^*)P^2, \nonumber \\
P^2&=&q^2+2q(1-q)P^2+(1-q)^2 P^4.
\label{eq:extinction1}
\end{eqnarray}
In the first line of \ref{eq:extinction1}, the first term stands for the probability of dying before doubling. The second term is the probability for the replicating agent to complete one round of doubling and subsequent death of the two populations that branched from the two copies. 

In the second line of \ref{eq:extinction1}, the first term represents the probability that both replicating agents die before completing their doubling rounds. The second term is the probability that one of the replicating agents will die before completing its doubling round. At the same time, the other will be able to complete its round of doubling but subsequently die alongside its progenitor. The third and last term is the probability that both replicating agents will complete their doubling rounds and then for all four agents to die. The probability that the replicating agent will die (P) is the same for all generations, leading to a self-consistent equation. 

Solving for $P$ In Eq. \ref{eq:extinction1}, we find two solutions, the trivial one $P=1$, i.e., certain extinction, and the non-trivial solution which yields $P=\frac{q}{1-q}$. Inserting the non-trivial $P$ in the first line of Eq. \ref{eq:extinction1}, we obtain the extinction probability as a function of $q$ and $q^*$:
\begin{eqnarray}
P^*=q*+(1-q^*)\frac{q^2}{(1-q)^2}.
\label{eq:extinction2}
\end{eqnarray}

The total extinction probability of $N$ uncorrelated replicating agents---$P^*_N$, is thus given by
\begin{eqnarray}
P^*_N= \Pi_j^N \left( q^*_j+(1-q^*_j ) \frac{q^2}{(1-q)^2} \right).
\label{eq:extinction3}
\end{eqnarray}
where $q^*_j$ is the probability of extinction of the $j^{th}$ replicating agent. 

If we further assume that all the probability distributions are exponential, the resulting equations simplify to the following equations
\begin{eqnarray}
q&=&q^*=\frac{1}{1+\frac{\mu}{\Gamma_D}}, \\
P&=&P^*=\begin{cases} \mbox{1} & \mu \le \Gamma_D  \\ 
\frac{\Gamma_D}{\mu} & \mbox{otherwise} \end{cases}\\
P^*_N&=&\left( \frac{\Gamma_D}{\mu} \right)^N.
\label{eq:extinction4}
\end{eqnarray}

To connect the above calculations to the extinction of a population of ribosomes in a cell, we recall that only a fraction $\alpha$ of the ribosomes are participating in producing copies of themselves, albeit indirectly, by synthesizing ribosomal proteins \cite{Terry1,Terry2,Terry3,ourPNAS}. 

Thus, the probability of extinction of a ribosome population is given by $P^*_N = \left( \frac{\Gamma_D}{\mu} \right) ^{N \alpha}$. We used this formula to calculate the dependence of the extinction probability of the ribosome population on the ribosome translation speed $v_{aa}$ and on the fraction of ribosomes in assembly $\phi$ via the equation we derived from our generalized growth law, Eq. \ref{eq:four} to obtain: $\alpha_i = (\mu(\phi_i+1)
+\Gamma_D) \frac{L_i}{v_{aa}}$. which leads to:
\begin{eqnarray}
P^*_N&=\left( \frac{\Gamma_D}{\mu} \right)^{N \alpha_i} = \left( \frac{\Gamma_D}{\mu} \right)^{N(\mu(\phi_i+1)
+\Gamma_D) \frac{L_i}{v_{aa}}}.
\label{eq:extinction5}
\end{eqnarray}
This equation was used to draw the solid green line in Figure \ref{fig:three}.

Finally, we compared this simplified model with an event-driven stochastic Monte Carlo simulation we devised. In short, the simulation consists of a finite population of ribosomes. Each ribosome can randomly translate with probability $\alpha_i$ the $i^{th}$ ribosomal protein mRNA. Each ribosome has a random lifetime derived from a shifted gamma probability distribution---although any other realistic distribution, i.e., zero measure at zero time and finite first and second moments, yields similar results, as we verified. We used an average lifetime of $\Gamma_D^{-1}=300$ min. The synthesis of a ribosomal protein that consists of $L_i$ amino acids was also drawn from a shifted gamma distribution with an average synthesis rate of $\frac{L_i}{v_{aa}}$ min. Each time a complete set of ribosomal proteins has completed synthesis, a new ribosome assembly process is initiated in the simulation. The assembly process ends after a random assembly duration with an average of $\tau_{SA}=2$ min with a shifted gamma distribution. Upon completion of the assembly process, the newly formed ribosome joins the pool of existing ribosomes that synthesize various proteins, including more ribosomal proteins. Hence, if there are enough ribosomes, the ribosome population increases exponentially over time. 

To empirically estimate the probability of extinction of the ribosome population, we begin the simulation with $5400$ ribosomes in total. For each point in the figure, i.e., for a given translation speed $v_{aa}$ we found the required allocation parameter $\alpha$ that yielded the same doubling time of $160$ min for all the points in the graph. We let each simulation run until the ribosome population became extinct or grew exponentially and crossed the arbitrary limit of $10$ times the initial number of ribosomes allocated to ribosomal protein synthesis. We repeat the simulation for $1024$ times to obtain the extinction probability estimate, plotted in Fig. \ref{fig:three}.

Perhaps surprisingly, while the theoretical model does not account for the intricacies of the ribosome assembly process, as seen in Fig. \ref{fig:three}, it agrees well with the simulation. 

\acknow{This research was supported by a grant from the United States-Israel Binational Science Foundation (BSF, Grant no. 2022763), Jerusalem, Israel, and the United States National Science Foundation (NSF) and by the Israel Science Foundation (Grant no. 776/19). We acknowledge Oren Raz and Fangwei Si for their useful comments on the manuscript}

\showacknow{} 



\end{document}